\documentstyle[11pt,paspconf,psfig]{article}

\newcommand{\kms}{km s$^{-1}$}
\newcommand{\lya}{Ly$\alpha$}

\begin{document}

\title{Constraints on the Unseen Galaxy Population from the \lya\ Forest}

\author{Kenneth M. Lanzetta and Hsiao-Wen Chen}
\affil{Department of Physics and Astronomy, State University of New York at
Stony Brook, Stony Brook, NY 11794--3800, U.S.A.}

\author{John K. Webb}
\affil{School of Physics, University of New South Wales, Sydney 2052, NSW,
AUSTRALIA}

\author{Xavier Barcons}
\affil{Instituto de F\'\i sica de Cantabria (Consejo Superior de 
Investigaciones Cient\'\i ficas---Universidad de Cantabria), Facultad de
Ciencias, 39005 Santander, SPAIN}

\begin{abstract}
Here we describe results of our attempt to determine what types of galaxies are
responsible for the \lya\ forest absorption systems, based on our ongoing
imaging and spectroscopic survey of faint galaxies in fields of HST
spectroscopic target QSOs.  Our primary conclusions are that the bulk of the
\lya\ forest arises in more or less normal galaxies (that span the normal range
of luminosity and morphology) and that any ``unseen'' low surface brightness
galaxies are unlikely to contribute significantly to the luminosity density of
the universe.
\end{abstract}

\keywords{galaxies, QSO absorption systems}

\section{Introduction}

  Our analysis of the relationship between galaxies and \lya\ absorption systems
at redshifts $z < 1$ has over the past few years led us to conclude that (1)
most galaxies possess extended gaseous envelopes of $\approx 160 \ h^{-1}$ kpc
radius and (2) many or most \lya\ absorption systems arise in extended gaseous
envelopes of galaxies (e.g.\ Lanzetta et al.\ 1995; Barcons et al.\ 1995; Chen
et al.\ 1998).  These conclusions bear directly, of course, on questions
concerning the nature and physical state of gaseous material at very large
galactocentric distances.  But what is of more immediate relevance to the topic
of this meeting is that these conclusions also imply that the ``forest'' of
\lya\ absorption lines that are routinely observed in the spectra of background
QSOs probe {\em galaxies} (rather than something else) to redshifts as large as
$z \approx 5$.

  The \lya\ resonance transition is an extraordinarily sensitive tracer of very
low column density material.  Ordinary spectra of ordinary QSOs can easily
detect neutral hydrogen column densities as low as $N \approx 5 \times 10^{13}$
cm$^{-2}$, which is many orders of magnitude below the column densities probed
by star light or 21 cm emission.  In this sense, the \lya\ forest represents a
more or less complete inventory of the baryonic constituents of the universe. 
Galaxies that are difficult to detect by ``ordinary'' means (i.e.\ by means of
star light or 21 cm emission) are easy to detect by means of \lya\ absorption
lines, so it remains to establish what portion of the \lya\ forest can be
attributed to ``normal'' galaxies that are represented by normal galaxy
luminosity functions in order to determine what portion of the \lya\ forest is
left over for ``unseen'' galaxies that are not represented by normal galaxy
luminosity functions.

  Here we describe the results of our attempt to determine what types of
galaxies are responsible for the \lya\ forest absorption systems, based on our
ongoing imaging and spectroscopic survey of faint galaxies in fields of Hubble
Space Telescope (HST) spectroscopic target QSOs.  Using new observations of the
galaxies of the survey, we have sought to establish just what factors play a
role in determining the gaseous extent of galaxies.  Our primary conclusions are
that the bulk of the \lya\ forest arises in more or less normal galaxies (that
span the normal range of luminosity and morphology) and that any ``unseen'' low
surface brightness galaxies are unlikely to contribute significantly to the
luminosity density of the universe.  More stringent conclusions along these
lines are within reach, requiring only further observations and analysis. 
Throughout we adopt a standard Friedmann cosmological model of deceleration
parameter $q_0 = 0.5$ and Hubble constant $H_0 = 100 \ h$ \kms\ Mpc$^{-1}$.

\section{Imaging and Spectroscopic Survey}

  Over the past several years, we have been conducting an ongoing imaging and
spectroscopic survey of faint galaxies in fields of HST spectroscopic target
QSOs.  The goal of the survey is to determine the gaseous extent of galaxies and
the origin of \lya\ absorption systems by directly comparing the redshifts of
galaxies and absorbers identified along common lines of sight.  The observations
have been and will be described elsewhere (e.g.\ Lanzetta  et al.\ 1995; Chen et
al.\ 1998), but in summary the observations consist of (1) optical images and
spectroscopy of objects in the fields of the QSOs, obtained with various
telescopes and from the literature, and (2) ultraviolet spectroscopy of the
QSOs, obtained with the HST using the Faint Object Spectrograph (FOS) and
accessed through the HST archive.  The optical images and spectroscopy are used
to identify and measure galaxy redshifts and impact parameters, and the
ultraviolet spectroscopy is used to identify and measure absorber redshifts and
equivalent widths.

  A total of 352 galaxies and 230 absorbers toward 24 fields are included into
the current analysis.  The galaxies and absorbers are ``matched'' or
``associated'' using quantitative criteria that are set by the galaxy--absorber
cross-correlation function $\xi_{\rm ga}(v,\rho)$, as it depends on the
line-of-sight velocity separation $v$ and the transverse impact parameter
separation $\rho$.  (Here we adopt the galaxy--absorber cross-correlation
function measured previously by Lanzetta et al.\ 1997 on the basis of 3126
galaxy and absorber pairs.)  In this way, ``physical'' galaxy and absorber pairs
are quantitatively distinguished from ``correlated'' and ``random'' galaxy and
absorber pairs.  Galaxies and absorbers within 3000 \kms\ of the QSOs are
excluded in order to focus the analysis on the ``intervening'' population.

\begin{figure}
\centerline{\psfig{file=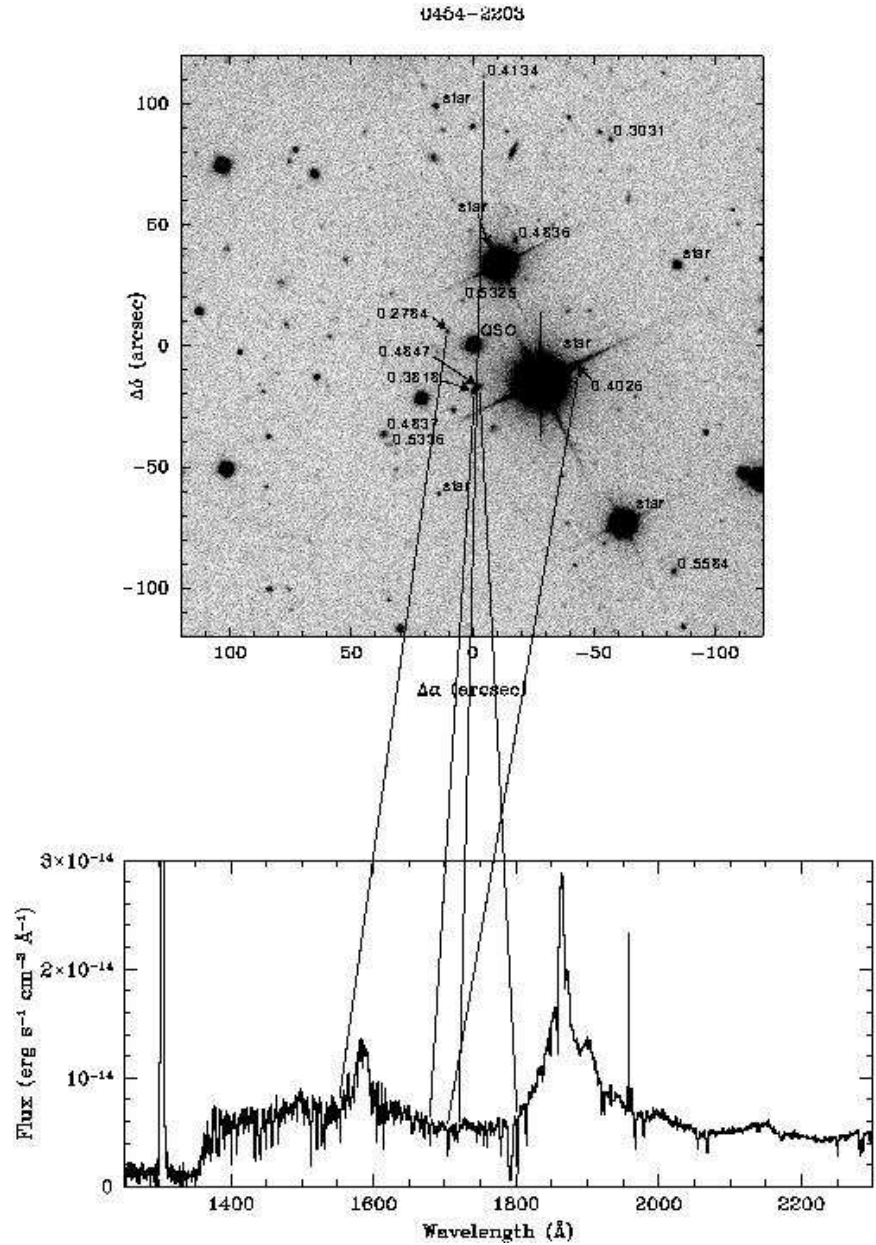,width=4.5in}}
\caption{Schematic illustration of results of the survey.  Top panel shows an
image of the file surrounding 0454$-$2203.  Bottom panel shows a spectrum of
0454$-$2203.  Redshifts of the various faint galaxies so far identified by the
survey are indicated in the top panel, and \lya\ absorption lines matched with
galaxies so far identified by the survey are indicated in the bottom panel.
Redshift of 0454$-$2203 is $z = 0.54$, and the most prominent emission line in
the spectrum is \lya.}
\end{figure}

  Results of the survey are illustrated schematically in Figure 1.  Figure
shows in the top panel an image of the field surrounding 0454$-$2203 and in the
bottom panel a spectrum of 0454$-$2203.  Redshifts of the various faint galaxies
so far identified by the survey are indicated in the top panel, and \lya\
absorption lines matched with galaxies so far identified by the survey are
indicated in the bottom panel.  All matched galaxy and absorber pairs indicated
in Figure 1 have cross-correlation amplitudes satisfying $\xi_{\rm ga}(v,\rho) >
1$.

\section{Gaseous Extent of Galaxies}

  One of the most striking results of the survey is that there exists a
distinctive anti-correlation between \lya\ absorption equivalent width $W$ and
galaxy impact parameter $\rho$.  In particular, galaxies at impact parameters
less than $\approx 160 \ h^{-1}$ kpc are {\em almost always} associated with
corresponding \lya\ absorption systems while galaxies at impact parameters
greater than $\approx 160 \ h^{-1}$ kpc are {\em almost never} associated with
corresponding \lya\ absorption systems.  The anti-correlation is statistically
highly significant and persists even when various subsamples (e.g.\ absorption
systems that exhibit metal absorption lines, or absorption systems that exhibit
very strong \lya\ absorption lines) are arbitrarily removed from the analysis.
On the basis of this result, we conclude that galaxies are surrounded by
extended gaseous envelopes of $\approx 160 \ h^{-1}$ kpc radius. The
anti-correlation between \lya\ absorption equivalent width and galaxy impact
parameter is shown in the top panel of Figure 2.

\begin{figure}
\centerline{\psfig{file=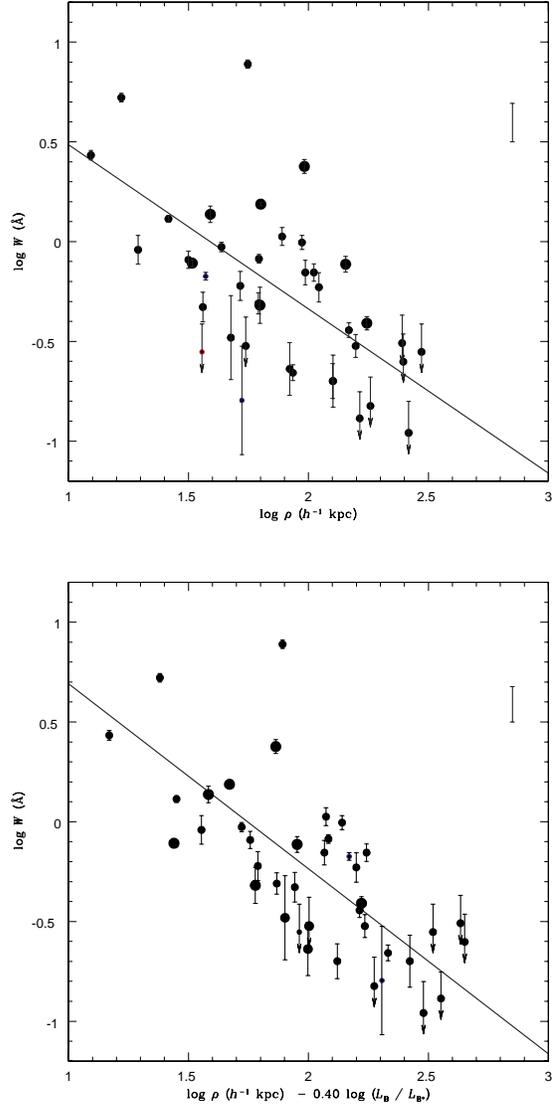,width=3.0in}}
\caption{{\em Top panel:}  Anti-correlation between \lya\ absorption equivalent
width $W$ and galaxy impact parameter $\rho$.  {\em Bottom panel:}
Anti-correlation between \lya\ absorption equivalent width $W$ and galaxy impact
parameter $\rho$ as ``corrected'' by the best-fit scaling with galaxy $B$-band
luminosity $L_B$.  Symbol size indicates galaxy luminosity (larger symbols
indicate larger luminosities) and symbol type indicates galaxy morphological
type (circles for elliptical and S0 galaxies, triangles for early-type spiral
galaxies, and squares for late-type spiral galaxies).  The error bars in the
upper right corners indicate ``cosmic'' scatter.  Both panels are based on
galaxies for which HST WFPC2 images have already been obtained.}
\end{figure}

  Yet it is clear from Figure 2 that the scatter about the mean relationship
between \lya\ absorption equivalent width and galaxy impact parameter is quite
substantial.  Evidently the amount of gas encountered along the line of sight
depends on other factors besides galaxy impact parameter, including perhaps
galaxy luminosity, size, or morphological type; the geometry of the impact
(e.g.\ if tenuous gas is distributed around galaxies in flattened disks rather
than in spherical halos); or disturbed morphologies or the presence of close
companions (e.g.\ if tenuous gas is distributed around galaxies as a result of
interactions).  To determine these other factors, we initiated a program to
obtain and analyze HST Wide Field Planetary Camera 2 (WFPC) images of galaxies
identified in the survey.  These observations were obtained (and are being
obtained) in HST Cycles 5 and 6.

  Using the WFPC2 images together with existing spectroscopic observations, we
measured properties of galaxies identified in the survey, including rest-frame
$B$-band luminosity $L_B$, effective radius $r_e$, average surface brightness
$\langle \mu \rangle$, disk-to-bulge ratio $D/B$, redshift $z$, and inclination
and orientation angles.  We then applied multivariate analysis techniques to
search for a ``fundamental surface'' in the multidimensional space that is
spanned by various combinations of the measurements.  lnitial results of the
analysis are described by Chen et al.\ (1998).

  The primary result of the analysis is that the amount of gas encountered
along the line of sight depends on the galaxy impact parameter and $B$-band
luminosity but does not depend strongly on the galaxy average surface
brightness, disk-to-bulge ratio, or redshift.  Spherical halos cannot be
distinguished from flattened disks on the basis of the current observations,
and there is no evidence that galaxy interactions play an important role in
distributing tenuous gas around galaxies in most cases.  These results are
presented in the bottom panel of Figure 2 and in Figure 3.  The bottom panel of
Figure 2 shows the anti-correlation between \lya\ absorption equivalent width
and galaxy impact as ``corrected'' by the best-fit scaling with galaxy $B$-band
luminosity.  Figure 3 shows the residuals with respect to the best-fit \lya\
absorption equivalent width versus galaxy impact parameter relation as
functions of galaxy $B$-band absolute magnitude $M_B$, redshift $\log (1 + z)$,
average surface brightness $\langle \mu \rangle$, and disk-to-bulge ratio $\log
D/B$.

\begin{figure}
\centerline{\psfig{file=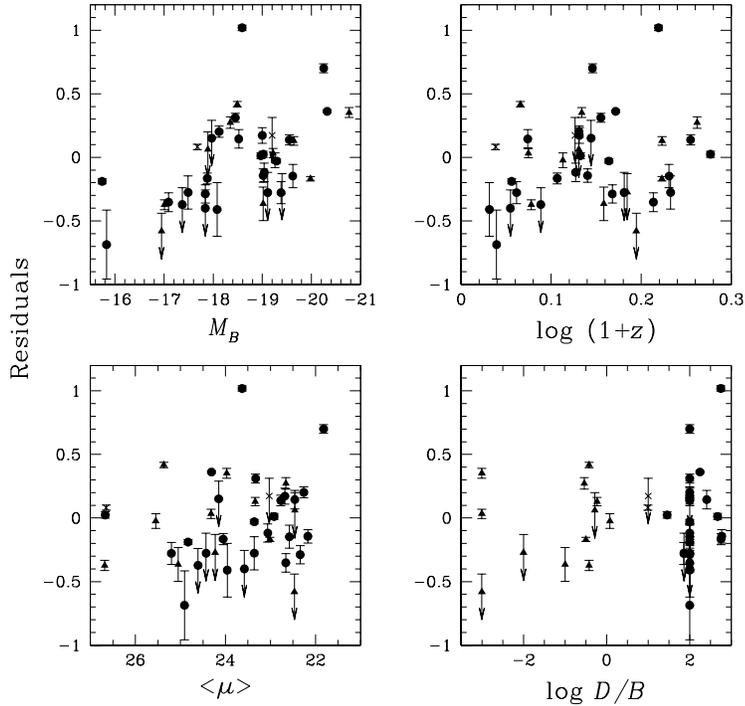,width=4.0in}}
\caption{Residuals with respect to the best-fit \lya\ absorption equivalent
width versus galaxy impact parameter relation as functions of galaxy $B$-band
absolute magnitude $M_B$, redshift $\log (1 + z)$, average surface brightness
$\langle \mu \rangle$, and disk-to-bulge ratio $\log D/B$.  Symbols are as for
Figure 2.}
\end{figure}

  We ascribe two especially important implications to the remarkably tight
anti-correlation between \lya\ absorption equivalent width and galaxy impact as
corrected by the best-fit scaling with galaxy $B$-band luminosity:  First, it
indicates that we have generally matched the appropriate galaxies with the
appropriate absorption systems.  Second, it indicates that galaxies are
surrounded by distinct gaseous ``envelopes'' and that these envelopes are
tightly associated with the individual galaxies, rather than loosely associated
with the ``large-scale environments'' of the individual galaxies, e.g.\ as
inter-group or inter-cluster gas.  The result that the amount of gas encountered
along the line of sight does not depend strongly on the galaxy average surface
brightness, disk-to-bulge ratio, or redshift apparently indicates that extended
gaseous envelopes are a common and generic feature of galaxies spanning a wide
range of luminosity and morphological type and therefore that the \lya\ forest
traces a representative portion of the galaxy population.

  The scaling relationship between galaxy gaseous radius $R$ and galaxy
$B$-band luminosity $L_B$ is well described by
\begin{equation}
\frac{R}{R_*} = \left( \frac{L_B}{L_{B_*}} \right) ^t.
\end{equation}
This relationship is analogous to the Holmberg relationship between galaxy
stellar radius and galaxy $B$-band luminosity.  Based on a sample that is
slightly larger than the sample analyzed by Chen et al.\ (1998), the best-fit
parameter estimates are (Chen et al.\ 1999, in preparation)
\begin{equation}
t = 0.40 \pm 0.09
\end{equation}
and
\begin{equation}
R_* = 190 \pm 34 \ h^{-1} \ {\rm kpc},
\end{equation}
which applies for \lya\ absorption equivalent widths satisfying $W > 0.32$ \AA.
The most important implication of the scaling relationship of equation (1) is
that it provides, for the first time, a means of quantitatively relating
statistical properties of \lya\ absorption systems to statistical properties of
galaxies.

\section{Constraints on the Unseen Galaxy Population}

  The predicted number density $n(z)$ of \lya\ absorption systems that arise in
extended gaseous envelopes of galaxies is
\begin{equation}
n(z) = \frac{c}{H_0} (1 + z) (1 + 2 q_0 z)^{-1/2} \int_{L_{B_{\rm min}}}^\infty
dL_B \Phi(L_B, z) \pi R^2(L_B),
\end{equation}
where $c$ is the speed of light, $H_0$ is the Hubble constant, $z$ is the
redshift $\Phi(L_B, z)$ is the galaxy luminosity function, $R(L_B)$ is the
galaxy gaseous radius, and $L_{B_{\rm min}}$ is the minimum galaxy luminosity
under consideration.  By adopting the {\em known} (from equation 1)
relationship between galaxy gaseous radius and galaxy $B$-band luminosity and a
{\em known} galaxy luminosity function, comparison of the {\em predicted} and
{\em observed} number densities of \lya\ absorption systems constrains
``unseen'' galaxies that are not represented by the galaxy luminosity function.

\begin{figure}
\centerline{\psfig{file=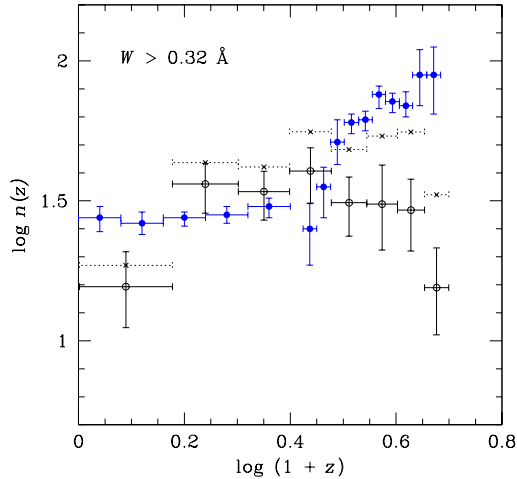,width=3.0in}}
\caption{Observed (solid circles) and predicted (open circles) number densities
of \lya\ absorption sytems.  Observed measurements are taken from Bechtold
(1994) and Weymann et al.\ (1998).  Crosses show predicted number density of
\lya\ absorption systems corrected for incompleteness at faint galaxy
luminosities.}
\end{figure}

  Results of this comparison are shown in Figure 4.  Figure 4 shows the
observed number density of \lya\ absorption systems (solid circles) and the
predicted number density of \lya\ absorption systems (open circles), based on a
galaxy luminosity function determined from photometric redshifts of galaxies in
the Hubble Deep Field (Fern\'andez-Soto, Lanzetta, \& Yahil 1998).  Figure 4
also shows the predicted number density of \lya\ absorption systems corrected
for incompleteness at faint galaxy luminosities (crosses).  The primary result
of Figure 4 is that to within measurement error {\em known} galaxies of {\em
known} gas cross sections can account for {\em all} \lya\ absorption systems at
redshifts $z < 2$ (and for most or all \lya\ absorption systems at higher
redshifts, after allowing for incompleteness at faint galaxy luminosities.)
This suggests that the ``unseen'' galaxy population produces at most a small
fraction of the \lya\ absorption systems.  For the scaling relationship of
equation (1) with $t = 0.4$, equation (4) indicates that the number density of
\lya\ absorption systems is roughly proportional to the $B$-band luminosity
density of the universe (multiplied by weakly redshift-dependent factor).  We
thus conclude that any ``unseen'' low surface brightness galaxies are unlikely
to contribute significantly the the luminosity density of the universe.

\acknowledgments

This research was supported by NASA grant NAGW--4422 and NSF grant
AST--9624216.

\end{document}